\documentclass{phb-proc4-auth}
\usepackage{graphicx}
\usepackage{amsmath}
\usepackage{amssymb}

\begin{document}

\begin{frontmatter}
\journal{SCES '04}
\title{Luttinger liquid versus charge density wave behaviour 
in the one-dimensional spinless fermion Holstein model}
\author[HGW]{H. Fehske}$^{,*}$,
\author[ER]{G. Wellein},
\author[ER]{G.~Hager},
\author[SY]{A. Wei{\ss}e},
\author[DD]{K. W. Becker},
\author[LA]{A.R. Bishop}
\corauth{Corresponding Author: 
%    Somewhere, MI 12345, USA.  Phone: (555) 555-5555 
%    Fax: (555) 555-7777, Email: JDoe@uol.edu\\
%Institut f\"ur Physik, 
%Universit\"at Greifswald,
%  D-17487 Greifswald, Germany.
  Phone: +49-3834-86-4760, Fax +49-3834-86-4701, 
  Email: fehske@physik.uni-greifswald.de}
\address[HGW]{Institut f\"ur Physik, 
Universit\"at Greifswald,
  D-17487 Greifswald, Germany}
\address[ER]{Regionales Rechenzentrum Erlangen, 
Universit\"at Erlangen-N\"urnberg,
  Germany}
\address[SY]{School of Physics, The University of New South Wales,
Sydney, NSW 2052, Australia}
\address[DD]{Institut f\"{u}r Theoretische Physik,
  TU Dresden, D-01062 Dresden, Germany}
\address[LA]{Theoretical Division, Los Alamos National Laboratory, Los
  Alamos, New Mexico 87545, USA}
\begin{abstract}
We discuss the nature of the different ground states 
of the half-filled Holstein model of spinless fermions in 1D. 
In the metallic regime we determine the renormalised effective 
coupling constant and the velocity of the charge excitations 
by a density-matrix renormalisation group (DMRG) finite-size scaling approach.
At low (high) phonon frequencies the Luttinger liquid is characterised
by an attractive (repulsive) effective interaction. 
In the charge-density wave Peierls-distorted state the 
charge structure factor scales to a finite 
value indicating long-range order. 
\end{abstract}
\end{frontmatter}
The challenge of understanding quantum phase transitions in 
novel quasi-1D materials  has stimulated intense work on 
microscopic models of interacting electrons and phonons such 
as the Holstein model of spinless fermions (HMSF) 
\begin{eqnarray}\label{model}
  H &=& -t\sum_i(c_i^{\dagger} c_{i+1}^{} +c_{i+1}^{\dagger} c_{i}^{})
+\omega_0\sum_i b_i^{\dagger} b_{i}^{}\nonumber\\
&&-g\omega_0\sum_i(b_i^{\dagger} + b_{i}^{})(n_i-\tfrac{1}{2})\,.
\end{eqnarray}
The HMSF describes tight-binding band electrons coupled locally to 
harmonic dispersionless optical phonons, 
where $t$, $\omega_0$, and $g$ denote the electronic transfer amplitude,
the phonon frequency, and the electron-phonon (EP) coupling constant, 
respectively.

Despite of its simplicity the HMSF is not exactly solvable  
and a wide range of numerical methods has been applied in the past to 
map out the ground-state phase diagram in the $g$-$\omega_0$-plane,  
in particular for the half-filled band case ($N_{el}=N/2$). There, the model  
most likely exhibits a transition from a Luttinger liquid (LL) to 
a charge-density wave (CDW) ground state above a critical EP coupling 
strength $g_c(\omega_0)>0$~\cite{BMH98}.

In this contribution we present large-scale DMRG calculations, 
providing unbiased results for the (non-universal) LL parameters
$u_\rho$, $K_{\rho}$, and the staggered charge structure factor
$S_c(\pi)$.

To leading order, the charge velocity $u_\rho$ and the 
correlation exponent $K_{\rho}$ might be obtained 
from a finite-size scaling of the of the ground-state energy 
of a finite system $E_0(N)$ with $N$ sites  
\begin{equation}\label{LLscaling1}
 \varepsilon_0(\infty)-(E_0(N)/N)=(\pi/3) (u_\rho/2)/N^{2}
\end{equation}
($\varepsilon_0(\infty)$ denotes the bulk ground-state energy density)
and the charge excitation gap 
\begin{equation}\label{LLscaling2}
\Delta_{ch}(N)=E^{(\pm1)}_0(N)-E_0(N)=\pi (u_\rho/2)/(NK_\rho) 
\end{equation}
(here $E^{\pm1}_0(N)$ is the ground-state 
energy with $\pm1$ fermions away from half-filling).
The LL scaling relations (2) and (3) were  
derived for the pure electronic spinless fermion model 
only~\cite{CV84}.
\begin{figure}[t]
\centering
  \includegraphics[width=.9\linewidth]{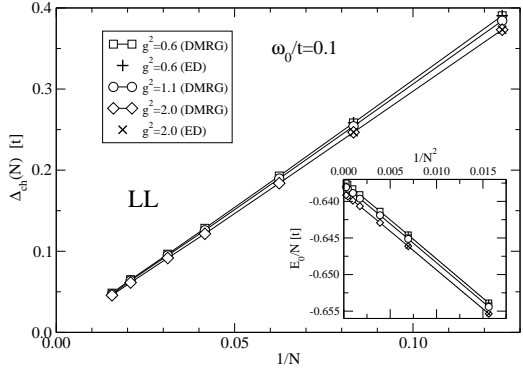}
  \caption{Finite-size scaling of the charge gap $\Delta_{ch}(N)$  
and the ground-state energy $E_0(N)$. Exact diagonalisation (ED) 
data is included for comparison.}\label{fig1}
\end{figure}
\setlength{\tabcolsep}{4mm}

Figure 1 demonstrates, exemplarily for the adiabatic regime, 
that they also hold for the case that a finite EP coupling is included.
The resulting LL parameters are given in Tab. 1.
Most notably, around $\omega_0/t\sim 1$, the LL phase splits 
in two different regimes: For small phonon frequencies the 
effective fermion-fermion interaction is {\it attractive}, 
while it is {\it repulsive} for large frequencies. 
In the latter region the kinetic energy is strongly reduced and 
the charge carriers behave like (small) polarons.
In between, there is a transition line $K_\rho=1$, where
the LL is made up  of (almost) non-interacting particles.      
The LL scaling breaks down just at $g_c(\omega/t)$, i.e. at the
transition to the CDW state. We found  $g_c^2 (\omega/t=0.1)\simeq 7.84$ 
and $g_c^2(\omega/t=10)\simeq 4.41 $~\cite{FHW00}.  
\begin{table}[h] 
\centering
\begin{tabular}{ccccc}\\ \hline
\rule[0mm]{0mm}{5mm} $g^2$&\multicolumn{2}{c}{$\omega_0/t=0.1$}&
\multicolumn{2}{c}{\mbox{$\omega_0/t=10.0$}}\\
 \rule[-3mm]{0mm}{6mm}  & $K_\rho$  & $u_\rho/2$  &  $K_\rho$    &  $u_\rho/2$\\ \hline
\rule[0mm]{0mm}{5mm} 0.6   & 1.031     &  $\sim 1$  &  $\sim 1$    &  0.617\\ 
\rule[0mm]{0mm}{1mm} 2.0   & 1.055     &  0.995    &   0.949 & 0.146\\
\rule[-3mm]{0mm}{6mm} 4.0   & 1.091     &  0.963    &   0.651 & 0.028 \\\hline\\
\end{tabular}
\caption{LL parameters 
at small and large phonon frequencies.}
\end{table}

Figure 2 proves the existence of the long-range ordered 
CDW phase above $g_c$. Here the charge structure factor   
\begin{equation}\label{scf}
S_c(\pi)=\frac{1}{N^2}\sum_{i,j}(-1)^j\langle (n_i-\tfrac{1}{2})
(n_{i+j}-\tfrac{1}{2})\rangle 
\end{equation}
unambiguously scales to a finite value in the thermodynamic limit
($N\to \infty$). 
Simultaneously $\Delta_{ch}(\infty)$ acquires a finite value.   
In contrast we have $S_c(\pi)\to 0$ in the metallic regime  ($g<g_c$). 
The CDW for strong EP coupling is connected to a Peierls distortion of the 
lattice, and can be classified as traditional band insulator 
and bipolaronic insulator in the strong-EP coupling adiabatic 
and anti-adiabatic regimes, respectively.

The emerging physical picture can be summarised 
in the schematic phase diagram shown in Fig.~3.
In the adiabatic limit ($\omega_0\to 0$) any finite EP coupling
causes a Peierls distortion. 
In the anti-adiabatic strong EP coupling limit ($\omega_0\to \infty$),
the HMSF can be mapped perturbatively onto the XXZ model
and the metal-insulator transition is consistent with a 
Kosterlitz-Thouless transition at $g_c^2(\infty)\simeq 4.88$.  

\begin{figure}[t]
  \includegraphics[width=\linewidth]{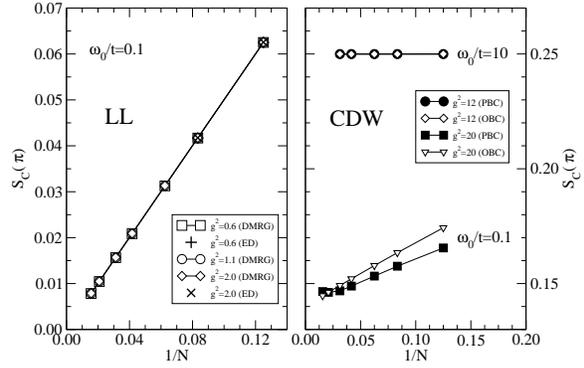}
  \caption{Scaling of the charge structure factor $S_c(\pi)$ using 
periodic (PBC) and open (OBC) boundary conditions.}\label{fig2}
\end{figure}
\begin{figure}[h]
\centering
 \includegraphics[width=\linewidth]{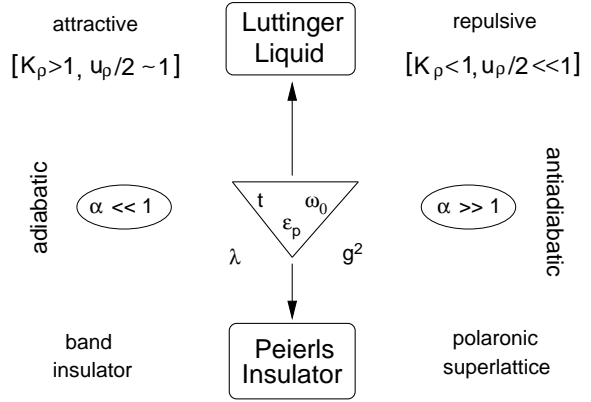}
  \caption{Phase diagram of the 1D half-filled spinless fermion 
Holstein model. Here $\alpha=\omega_0/t$, $\varepsilon_p=\omega_0g^2$, 
and $\lambda=\varepsilon_p/2t$.}\label{fig3}
\end{figure}

Work was supported by DFG under SPP 1073 and SFB 463, KONWIHR, HLRN Berlin,
and the  US DOE.

\end{document}